\documentclass[12pt]{article}
\usepackage{longtable}
\newcommand{\xxx}[1]{ [#1]}
\newcommand{\mysection}[1]{\section{#1}
   \hspace{0.8cm}\setcounter{equation}{0}}
\renewcommand{\theequation}{\arabic{section}.\arabic{equation}}
\newcommand{\myappendix}{\appendix
   \renewcommand{\theequation}{\Alph{section}.\arabic{equation}}
   \vspace{30pt} \noindent {\Large \bf Appendices}}
\newlength{\dummysp}
\newcommand{\beq}{\begin{equation}}
\newcommand{\eeq}{\end{equation}}
\newcommand{\ben}{\begin{enumerate}}
\newcommand{\een}{\end{enumerate}}

\newcommand{\mtxt}[1]{\mathop{\hbox{{\small #1}}}\nolimits}

\newcommand{\stxt}[1]{\mathop{\hbox{{\scriptsize #1}}}\nolimits}
\newcommand{\bbar}[1]{{\overline{#1}}}

\newcommand{\half}{{1 \over 2}}

\newcommand{\beqa}{\begin{eqnarray}}
\newcommand{\eeqa}{\end{eqnarray}}

\newcommand{\mod}{{\; \mtxt{mod} \; }}

\newcommand{\Zbf}{{{\bf Z}}}

\newcommand{\gappeq}{\mathrel{\rlap {\raise.5ex\hbox{$>$}}
{\lower.5ex\hbox{$\sim$}}}}
\newcommand{\lappeq}{\mathrel{\rlap{\raise.5ex\hbox{$<$}}
{\lower.5ex\hbox{$\sim$}}}}
\newcommand{\myref}[1]{(\ref{#1})}

\newcommand{\ite}[1]{{\it #1 \hspace{2pt}}}

\newcommand{\so}{$SO(32)$}
\newcommand{\ti}{\times}
\newcommand{\GNA}{G_{\stxt{NA}}}
\newcommand{\mo}{(-1)}

\newcommand{\un}[1]{\underline{#1}}
\newcommand{\eetee}{$E_8 \ti E_8$}
\newcommand{\tspt}{${\rm spin}(32)/Z_2$}
\newcommand{\spl}{\sigma_+}
\newcommand{\tK}{{\tilde K}}
\def\[{\left[}
\def\]{\right]}
\def\({\left(}
\def\){\right)}

\begin{document}

\begin{titlepage}

\baselineskip=14pt

\renewcommand{\thefootnote}{\fnsymbol{footnote}}

\hfill    hep-th/0301232

\hfill    Mar.~28, 2003

\vspace{10pt}

\begin{center}
{ \bf \Large $Z_3$ orbifolds of the $SO(32)$ heterotic string:
\\  \vskip 5pt
1-Wilson-line embeddings}
\end{center}

\begin{center}
{\sl Joel Giedt}\footnote{E-Mail: {\tt giedt@physics.utoronto.ca}}

\end{center}

\begin{center}

{\it Department of Physics, University of Toronto, \\
60 St. George St., Toronto, ON, M5S 1A7, Canada}

\end{center}

\begin{center}
{\bf Abstract}
\end{center}

We consider compactification of the $SO(32)$
heterotic string on a 6-dimensional $Z_3$
orbifold with one discrete Wilson line.
A complete set of all possible embeddings is given,
159 in all.
The unbroken subgroups of $SO(32)$ are tabulated.
The extended gauge symmetry $SU(3)^3$, recently
discussed by J.~E.~Kim [hep-th/0301177] for semi-realistic
$E_8 \ti E_8$ heterotic string models,
occurs for several embeddings,
as well as other groups that may be of interest
in unified string models.
The extent to which extra gauge
group factors can be hidden is discussed and
compared to the $E_8 \ti E_8$ case.
Along flat directions where an effective
hidden sector exists, the embeddings described here provide
for hidden gauge groups that are
not possible in the $E_8 \ti E_8$ heterotic string.

\vfill

\end{titlepage}

\renewcommand{\thefootnote}{\arabic{footnote}}
\setcounter{page}{1}
\setcounter{footnote}{0}

\baselineskip=14pt

\mysection{Introduction}
One of the main motivations for starting
with the \eetee\ heterotic string \cite{GHMR85} in semi-realistic
applications is that the second $E_8$ factor
would {\it appear} to provide a natural source of
a hidden sector in which to break supersymmetry
by, say, gaugino condensation~\cite{gcd}.  Indeed, in
a field theoretic dimensional reduction of the
10-dimensional theory to 4 dimensions, one finds
only Planck mass suppressed operators communicating
between the sets of gauge-charged fields coming
from the two $E_8$'s.

However, when the underlying string
theory is formulated with 6 dimensions compactified
on an orbifold~\cite{DHVW85,DHVW86}, quantum consistency of the 2-dimensional
conformal field theory necessitates the
addition of twisted sector states.
These are states that would not be present in a field theoretic
dimensional reduction of the original 10-dimensional
theory.  Quite generally, these twisted states
are simultaneously charged under subgroups of both $E_8$
factors~\cite{Gie02a}.  Thus these states can mediate supersymmetry
breaking through gauge interactions and the so-called
hidden sector is no longer hidden. Typically
one overcomes this difficulty by breaking the
gauge interactions (most often extra $U(1)$'s and $SU(2)$'s)
at or near the string scale so that their interactions
are more or less Planck scale suppressed.  This
requires a careful choice of flat direction;
for examples see~\cite{flc}.

Against this backdrop it is worth reconsidering
the aversion to the \so\ heterotic string, since: (i)
with an appropriate orbifold embedding chiral
representations occur, and (ii) extra gauge
symmetries can be hidden the same way that they
are hidden in \eetee\ heterotic orbifolds.  It will be
seen from the gauge groups obtained below that most
embeddings give rise to a product of nonabelian
simple groups and some $U(1)$'s.  Generically representations will
arise that ``feel'' any pair of factors.  However,
if we can choose a flat direction that renders these
states supermassive (say, near the string scale),
then it would seem that we can manufacture an effectively
hidden sector that is comparable to what may be
obtained from a 4-dimensional \eetee\ heterotic
string construction via orbifold compactification.
Indeed, it is generic in such constructions that
many states do decouple near the string scale,
due to the presence of an anomalous $U(1)$,
since the anomaly is cancelled by a counterterm
that induces a Fayet-Iliopoulos term \cite{FIU}.
The point is that in either case, \eetee\ or \so,
there exist difficulties hiding the hidden sector
and specific assumptions regarding flat directions
must be made \cite{dmstab}.

It should be remarked however, that we may not want
to hide extra gauge groups.  Rather, one may have
in mind gauge mediation scenarios (see \cite{GM}
for a review, and the extensive references
therein) with exotic messenger
states that feel both the strong dynamics of the
sector that breaks supersymmetry and the ordinary
gauge symmetries of the observable world.  In this
case the larger messenger representations that
are possible in the $SO(32)$ string could be advantageous.\footnote{In
the \eetee\ heterotic string only twisted states can ``talk''
between the two $E_8$'s.  The twisted
states must have smaller \eetee\ root torus winding numbers
and hence tend to be in smaller representations
of the gauge group.}

In either event, model builders might wish to enlarge
their vistas by considering orbifolds of the $SO(32)$
heterotic string.  In this regard it is useful to
have some idea for a good starting point.  Selecting
the string scale gauge symmetry is certainly an important
first step, and it is therefore worth knowing what is
possible and how to get it.  With this goal in mind,
in the present article we enumerate all inequivalent embeddings
with 1 discrete Wilson line for the 6-dimensional $Z_3$ orbifold of
the \so\ heterotic string.  Thus we perform the
analogue of previous calculations in the \eetee\
heterotic string; most notably those in \cite{IMNQ,CMM89,Gie01b}.

Models that contain unbroken gauge groups
such as $SU(5)^2$ are starting points for higher
affine level models.  It has been described in \cite{AFIU95}
for instance how the presence of $(5, \bar 5)
+ (\bar 5, 5)$ representations can be used to break to the
diagonal subgroup, which is realized at level 2,
with the requisite Higgses to break further to
the Standard Model gauge group.\footnote{For other works on
how higher affine level models can be constructed
and how they might afford semi-realistic
string GUTs, see Refs.~\cite{sgut:o,sgut:d,sgut:kt}.}
However, the authors of \cite{AFIU95}
have also concluded that for the $Z_3$ orbifolds
it is not possible to get a chiral spectrum for the resulting level $k=2$
$SU(5)$ GUT.  They argue that this is because the net number of
fermion generations vanishes in constructions
with the requisite $(5, \bar 5) + (\bar 5, 5)$ representations.
After some elementary calculations,
we have reached the same conclusion.  Therefore,
although some embeddings here do have $SU(5)^2$
and the requisite representations to obtain
a level $k=2$ diagonal subgroup with adjoint Higgses,
they are not viable routes to semi-realistic string GUTs, and one
should look elsewhere.

Another sort of ``unified'' model has recently been studied
by J.~E.~Kim for the \eetee\ heterotic $Z_3$ orbifold \cite{Kim03}.
This is the group $SU(3)^3$.  The advantages of this
approach over standard-like constructions are
notable (see \cite{Kim03} for further
details).  We would like to point out that several
of the embeddings listed here also give $SU(3)^3$ in
the surviving group; we leave as a topic for further
investigation the phenomenology of these models,
which should share in the advantages noted in \cite{Kim03}.

In Section 2 we describe the embeddings of the orbifold
action into the sixteen internal bosons responsible for
gauge symmetry in the low-energy theory.  In Section 3
we discuss equivalence relations that are used to reduce
the number of embeddings.  In Section 4 we summarize the
key elements used in constructing all consistent embeddings.
In Section 5 we address the various possible Wilson
lines and our results.  In Section 6 we state our conclusions.
In Appendix A we discuss an important set of equivalence
relations.  In Appendix B we present tables of the
159 embeddings found here.

\mysection{Abelian $Z_3$ embeddings}
The sixteen internal left-moving bosons $X^I(\spl)$---the gauge
degrees of freedom---are compactified on the \tspt\
lattice, which we will denote by $\Lambda$.  This
lattice consists of all 16-vectors of the form
\beq
(n_1,\ldots,n_{16}), \qquad
(n_1 + \half,\ldots,n_{16}+\half),
\label{vfb}
\eeq
subject to the constraints $n_I \in \Zbf$ and
$\sum_I n_I  = 0 \mod 2$.  We remind the reader
that ${\rm spin}(32)$ is the covering group for \so.
The \so\ roots are
\beq
(\un{\pm 1, \pm 1, 0^{14}}).
\label{oiy}
\eeq
Here, signs are not correlated.  Underlining
indicates all permulations are to be taken.
The ``exponent'' indicates that the entry is
repeated 14 times.  Analogous notations will
be used below.

The $Z_3$ orbifold is obtained as the quotient
of the 6-dimensional $SU(3)^3$ root torus
by simulataneous $2 \pi/3$ rotations in each of
the three complex planes labeled by
complex coordinates $z^i \; (i=1,3,5)$.  This {\it twist} in
the 6-dimensional compact space is embedded into
internal gauge degrees of freedom in an abelian
manner---through a shift:
\beq
z^i \to e^{2\pi i/3} z^i
\Rightarrow X^I(\spl) \to X^I(\spl) + \pi V^I.
\eeq
In addition we allow for the possibility of {\it
discrete Wilson lines} $a_i \; (i=1,3,5)$ which embed translations
in the 6-dimensional compact space.  It suffices
to specify this embedding for three such shifts,
due to constraints arising from the torus construction:
\beq
z^i \to z^i + 1 \Rightarrow
X^I(\spl) \to X^I(\spl) + \pi a_i^I \quad \forall \; i=1,3,5.
\eeq
The four embedding vectors are subject to consistency
relations that follow from world-sheet modular
invariance:
\beq
3V,3a_i \in \Lambda, \quad
3V^2 \in \Zbf, \quad 3a_i^2 \in \Zbf, \quad
3V\cdot a_i \in \Zbf, \quad 3 a_i \cdot a_j \in \Zbf.
\label{ccd}
\eeq
An infinite number of solutions to these conditions
exist.  Fortunately only a finite number of
inequivalent possibilities are contained in this
set, due to equivalence relations that we discuss
in Section \ref{eqr}.

The massless gauge-charged states are
characterized by 16-dimensional winding vectors.
In the untwisted sector we have for states
with nontrivial weights with respect to the
Cartan subalgebra:
\beq
K \in \Lambda, \qquad K^2 = 2.
\eeq
The Wilson lines enforce a projection on these
states.  Only those that satisfy
\beq
a_i \cdot K \in \Zbf  \quad \forall \; i=1,3,5
\label{jjer}
\eeq
survive.  Those that do survive fall into three
categories:
\beq
3 V \cdot K = \left\{
\begin{array}{cc}
0 \mod 3 & {\rm gauge} \\
1 \mod 3 & {\rm matter} \\
-1 \mod 3 & {\rm antimatter}
\end{array}
\right.
\label{hyre}
\eeq
In truth this is a further projection onto states
with differing right-moving quantum numbers.

For the twisted states, corresponding to string states
with nontrivial monodromy, we have weights $\tK$ which
satisfy
\beq
\tK^2 = {4 \over 3} - 2N_L, \qquad
\tK = K + V + \sum_{i=1,3,5} n_i a_i, \qquad
K \in \Lambda.
\label{tms}
\eeq
If left-moving oscillators are excited
in the 6-dimensional compact space, we can
have $N_L = 1/3$ or $2/3$.  (However, $N_L =2/3$
only has a solution to \myref{tms} if the
embedding is equivalent to the trivial one;
i.e., $V=a_i=0$.)
The integers $n_i = 0,\pm 1$ label fixed
point locations in each of the 3 complex planes.
Each twisted state is labeled by a triple
$(n_1,n_3,n_5)$.  Note that \myref{ccd}
implies $3 \tK \in \Lambda$.

\mysection{Equivalence relations}
\label{eqr}
The equivalence relations are essentially those
already alluded to in \cite{DHVW86} and discussed
in detail in \cite{CMM89} for $Z_3$ orbifolds
of the \eetee\ heterotic string.

\ite{Lattice group equivalence.}
This merely states that $V \to V + K$
and $a_i \to a_i + K_i$ yield equivalent
embeddings for any choice $K,K_i \in \Lambda$.

\ite{Weyl group equivalence.}
This states that for any \so\ root $e$ (given
in \myref{oiy}) the simultaneous Weyl reflection
\beq
V \to V - (V \cdot e) e, \qquad
a_1 \to a_1 - (a_i \cdot e) e,
\eeq
gives an equivalent embedding.
It is easy to check that the Weyl group here consists
of permutation of entries, pair-wise sign flips, and
compositions of these operations.  This is a considerable
simplification over the \eetee\ case where half-integral
roots exist that lead to a more complicated Weyl group.
(The half-integral \so\ weights are not roots.)

\ite{Fixed point label equivalences.}
First, we have $a_i \to -a_i$ for any of the Wilson
lines.  This is just a relabeling $n_i \to -n_i$
of the fixed points.  Second, we have $V \to V \pm a_i$
for any choice $i=1,3,5$.  This is a relabeling
$n_i \to n_i \mp 1$ of the fixed points.  For instance
the twisted sector winding vector is rewritten
\beq
\tK = K + V + \sum_j n_j a_j =
K + (V \pm a_i) + (n_j \mp 1) a_i
+ \sum_{j \not= i} n_j a_j
\eeq
to display that this is nothing but a relabeling
of fixed points, keeping in mind
$n_i \simeq n_i \mod 3$.  The complete twisted sector
spectrum will be the same; only the labeling will be
different.  It is easy to check that the projections
\myref{hyre} in the untwisted sector are unchanged,
due to~\myref{jjer}.

\mysection{Building blocks}
Recall from \myref{ccd} that the
embedding vectors $V,a_i$ must satisfy
$3V \in \Lambda,3a_i \in \Lambda$.  Taking
into account \myref{vfb}, the entries of
$3V,3a_i$ are either all integral or all
half-integral.  We easily
restrict to integral weight lattice vectors
$3V,3a_i$ using the lattice group equivalence;
simply add $3K$ where $K$ is any half-integral
lattice vector.  Repeatedly adding $3e_j$'s, where $e_j$'s are
the roots in \myref{oiy}, allows us to lower
the magnitudes of entries of $3V,3a_i$ until no entry
has magnitude greater than 2.
We can restrict to no more than one $\pm 2$
appearing in $3V,3a_i$ using the lattice group
equivalence:  addition of some $3 e_j$, where $e_j$ is
one of the roots \myref{oiy}, can be used to
eliminate any pair of $\pm 2$'s.
The self-consistency constraints in \myref{ccd},
which we find it convenient to write
\beq
(3V)^2 = 0 \mod 3, \qquad (3a_i)^2 = 0 \mod 3,
\eeq
provide a further restriction, and since $K \in \Lambda$
implies $K^2$ even, we only get {\it even} multiples of 3.
Thus,
\beq
(3V)^2 = 0 \mod 6, \qquad (3a_i)^2 = 0 \mod 6.
\eeq
Therefore we find that $3V,3a_i$ can only
belong to the set
\beq
\{ (0^{16}), \quad (1^6,0^{10}), \quad (1^{12},0^4), \quad
(2,1^2,0^{13}), \quad (2,1^8,0^7), \quad (2,1^{14},0) \} ,
\label{jat}
\eeq
up to ordering and sign permutations.  It follows
that these 6 vectors form the basis of all subsequent analysis.

We fix ordering and signs for the twist embeddings using
the Weyl group equivalence.  Then the
inequivalent twist embeddings $3V$
together with their unbroken subgroups of $SO(32)$
and untwisted matter are given in Table \ref{tbt}.
We find it convenient in what follows to concentrate
on how Wilson lines further break the gauge
group in the two distinct parts of the 16-dimensional
space, emphasized by the placement of a semicolon
in the entries for $3V$ in Table \ref{tbt}.  The
first subspace, where $3V^I \not= 0$, we will refer
to as the {\it nonzero sector.}  The second
subspace, where $3V^I = 0$, we will refer to
as the {\it zero sector.}  These are not to be
confused with the usual sectors of the Hilbert
space of the underlying conformal field theory.

\begin{table}
\begin{center}
\begin{tabular}{ccc}
$3V$ & $G$ & Untw. matter \\ \hline
$(0^{16})$ & $SO(32)$ & none \\
$(1^6;0^{10})$ & $SU(6) \ti SO(20) \ti U(1)$
&  $3[(\bbar{15},1) + (6,20)]$  \\
$(1^{12};0^{4})$ & $SU(12) \ti SO(8) \ti U(1)$
& $3[(\bbar{66},1) + (12,8_v)]$ \\
$(-2,1^2;0^{13})$ & $SU(3) \ti SO(26) \ti U(1)$
& $3[(\bbar{3},1) + (3,26)]$ \\
$(-2,1^8;0^{7})$ & $SU(9) \ti SO(14) \ti U(1)$
& $3[(\bbar{36},1) + (9,14)]$ \\
$(-2,1^{14};0)$ & $SU(15) \ti U(1)^2$
& $3[(\bbar{105}) + 2(15)]$ \\
\hline
\end{tabular}
\end{center}
\caption{Inequivalent twist embeddings.}
\label{tbt}
\end{table}

\mysection{Wilson line embeddings}
Inequivalent Wilson lines $3a_1$ are obtained by taking
sign and ordering permutations of the six vectors \myref{jat}
subject to the $V \cdot a_1$ constraint in \myref{ccd},
which we find it convenient to write
\beq
3V \cdot 3a_1 = 0 \mod 3 .
\label{yut}
\eeq
In the case of $V=0$ we have only the six possibilities
listed in Table \ref{tb1}.

For the effects of Wilson lines when $V \not= 0$ it
is convenient to focus on the nonzero sector and the
zero sector separately.  Condition \myref{yut} constrains
the entries of $a_1$ in the nonzero sector, whereas
it places no constraint on the entries of $a_1$ in the
zero sector.  As an example we consider $3V=(1^6;0^{10})$
in some detail.  For the other cases we merely state
our results, except as regards some none-too-apparent
equivalences.

\begin{table}
$$
\begin{array}{cc}
3(a_1^1,\ldots,a_1^6) & SU(6) \; {\rm Subgroup} \\ \hline
(0^6), (1^6), (-2,1^5) & SU(6) \\
(1,-1,0^4), (1^4,-1,0), (-2,-1,0^4), (-2,1^3,-1,0)
& SU(4) \times U(1)^2  \\
(1^3,0^3), (1^3,\mo^3), (-2,1^2,\mo^3), (-2,1^2,0^3)
& SU(3)^2 \times U(1)^1  \\
(1^2,\mo^2,0^2), (-2,1,\mo^2,0^2)
& SU(2)^3 \times U(1)^2 \\ \hline
\end{array}
$$
\caption{Entries in the nonzero sector for $3V=(1^6;0^{10})$.}
\label{tba}
\end{table}

In the nonzero sector the possibilities that exist for the
corresponding 6 entries of $3a_1$ are given in Table \ref{tba}.
We have indicated how the $SU(6)$ factor surviving
$V$ is broken by the Wilson line.  Equivalence relations
have been used to eliminate obvious redundancies.  In
an appendix we show that embeddings with $(-2,\mo^4,0)$ are
equivalent to embeddings with $(-2,1^3,-1,0)$.  This
is why we have dropped the former possibility.
For the zero sector we have the set
\beq
(1^9,-1) \quad {\rm and} \quad (1^n,0^{10-n}) \quad
n=0,1,\ldots,10 .
\eeq
Note that the sign in the first vector cannot
be removed by the pairwise sign flips without
disturbing $V$.  In all other cases signs can
be removed in the zero sector.  In the case where
we have for $a_1$ entries $(0^6)$ in the nonzero sector
of $V$, we must
also include the possibilities of
\beq
(-2,1^2,0^7) \quad {\rm and} \quad (-2,1^8,0)
\eeq
in the zero sector of $a_1$ since,
lacking $\pm 1$'s in the nonzero sector, we cannot
push the $-2$ into the nonzero sector using
lattice group equivalence.
That is, in the other cases we can eliminate an entry of $-2$
from the zero sector using the lattice group
equivalence $3a_1 \to 3a_1 + 3K$ where
$3K = (\underline{\pm 3,0^5};3,0^9)$.

It is simple to determine the possibilities that
are consistent with a given choice for the nonzero
sector, by comparing to the vectors in \myref{jat}
and requiring that the total $3a_1$ be one of these
up to sign and ordering permutations.  For example
if we choose $(1,-1,0^4)$ in the nonzero sector
then we have only
\beq
(1,-1,0^4) \oplus \{ (1^4,0^6), (1^9,\pm 1) \} ,
\eeq
corresponding the second and third vectors in \myref{jat}.
These are Embeddings 2.5-2.7 of Table~\ref{tb2}.
As another example if we choose $(-2,1^2,\mo^3)$ in the
nonzero sector then we have only
\beq
(-2,1^2,\mo^3) \oplus \{ (1^3,0^7), (1^9,0) \},
\eeq
corresponding to the last two vectors in \myref{jat}.
These are Embeddings 2.21 and 2.22 of Table~\ref{tb2}.

The possibilities for the zero sector lead
to breakings of the $SO(20)$ that survives
$V$.  These are given in Table \ref{tbb}.
Of course they are correlated to what
occurs in the nonzero sector.  Thus we obtain
as a complete list of consistent embeddings
for $3V=(1^6;0^{10})$ the entries of Table \ref{tb2}.
We also list the nonabelian part $\GNA$ of the surviving
gauge group $G$.  One should add as many $U(1)$'s as
are needed to have $G$ a rank 16 group.

\begin{table}
$$
\begin{array}{cc}
3(a_1^7,\ldots,a_1^{16}) & SO(20) \; {\rm Subgroup} \\ \hline
(0^{10}) & SO(20) \\
(1,0^9) & SO(18) \times U(1) \\
(1^{10-n},0^n) \; n=2,\ldots,8 &
SU(10-n) \times SO(2n) \times U(1) \\
(1^9,0), (2,\mo^8,0) &
SU(9) \times U(1)^2 \\
(1^9,\pm 1) & SU(10) \times U(1) \\
(-2,1^2,0^7) & SU(3) \times SO(14) \times U(1) \\ \hline
\end{array}
$$
\caption{Entries in the zero sector for $3V=(1^6;0^{10})$.}
\label{tbb}
\end{table}

Including the cases with $a_1=0$, it can be seen
from Tables \ref{tb1}-\ref{tb6} that we have a total
of 159 different embeddings when one discrete Wilson line is
permitted.  Although it is plausible some redundancy may
yet exist, most of it has been removed using the
equivalence relations described above.  Certainly
the list is complete, meaning that any consistent
embedding with one discrete Wilson line is equivalent
to one of the embeddings presented here.

\mysection{Conclusions}
We have argued that the \so\ heterotic string provides
an interesting starting point for semi-realistic
string phenomenology.  We have worked out
a complete list of all embeddings with one discrete
Wilson line for the symmetric $Z_3$ orbifold.
To our knowledge this calculation has not been
presented previously in the literature.
We have addressed most of the equivalences that
relate embeddings.  Some of the less obvious
equivalences have been described in detail
in an appendix.

We have commented briefly on the prospects for
obtaining a hidden sector and for softly broken
supersymmetry via gaugino condensation in this sector.
It is our conclusion that there is every reason
to believe that some of the models described here
will be viable in this respect subject to an
appropriate choice of flat direction.  We leave
explicit explorations of this conjecture for
future investigation.

Many avenues for future research present themselves.
One interesting possibility has to do with the
phenomenology of the $SU(3)^3$ embeddings, along the lines of \cite{Kim03}.
Research in this direction is in progress and
we hope to report on it shortly.
Another issue worth exploration is flat directions
that might lead to a level $k=2$ $SU(5)$ GUT
for the embeddings that contain $SU(5)^2$ as a proper subgroup,
such as those embeddings with $SU(6)^2$.
In this case, an important question is whether
or not it is possible to obtain a chiral spectrum
with respect to the diagonal subgroup.
A final issue we would like to mention
is whether or not gauge mediation is viable
for any of these models.

It is our hope that we have convinced the reader
that the \so\ heterotic string can provide
intriguing unified models.  Further research
on the phenomenological possibilities when
this is taken as the starting point would certainly
be a welcome supplement to what is a comparatively
sparse examination in the existing literature.

\vspace{20pt}

\noindent
{\bf \Large Acknowledgement}

\vspace{5pt}

\noindent
This work was supported by the National Science and
Engineering Research Council of Canada.

\myappendix

\mysection{Technical aspects of embedding equivalences}
In this appendix we address some of the more technical
details in uncovering the equivalence between different
embeddings.

One type of equivalence is particularly important,
since it is easy to overlook, but removes a good
deal of redundancy.  It is used on the nonzero sector
embeddings with a $-2$.  We show the intermediate step
which sketches out the proof of the equivalence:
\beq
(-2,1^{m+2},\mo^{3n+m}) \simeq (1^{m+3},2,\mo^{3n+m-1})
\simeq (-2,1^{3n+m-1},\mo^{m+3}).
\label{lur}
\eeq
In the first step we add a pair of $3$'s using
lattice equivalence.  In the second step we send
$3a_1 \to -3a_1$.  Whatever signs we had in the
zero sector are reversed when this is done.
In many cases the orginal signs in the zero
sector can be restored by pairwise sign flips.
In those cases where this is not true, we can always
obtain $(1^p,\pm 1)$ in the nonzero sector,
again by pairwise sign flips.  But both possibilities
are included in our tables, so there is nothing lost
by using the equivalence \myref{lur}.

For example, an equivalence of the form
\myref{lur} exists in the embeddings with
$3V=(1^6;0^{10})$.  In the nonzero sector
we have:
\beq
(-2,\mo^4,0) \simeq (-2,1^3,-1,0).
\eeq
It is for this reason that $(-2,\mo^4,0)$ does
not appear in our tables.

As a second example, consider the embeddings
with $3V=(1^{12};0^4)$.  In this case when there
is a $-2$ present in the nonzero sector part of
$3a_1$, we have the range $3V\cdot 3a_1=-12,-9,\ldots,9$.
However, equivalences of type \myref{lur}
need to be accounted for.  It is not too hard
to check that:
\ben
\item[(i)]
all cases of $3V\cdot 3a_1=-12$
are equivalent to one of the cases of $3V\cdot 3a_1=6$;
\item[(ii)]
all cases of $3V\cdot 3a_1=-9$
are equivalent to one of the cases of $3V\cdot 3a_1=3$;
\item[(iii)]
all cases of $3V\cdot 3a_1=-6$
are equivalent to one of the cases of $3V\cdot 3a_1=0$;
\een
Then we can restrict to $3V\cdot 3a_1=-3,0,\ldots,9$.
It is easy to check that the possibilities are
those listed in Table \ref{tb3}.

\newpage

\mysection{Embedding Tables}

\begin{longtable}{ccc}
\caption{$3V=(0^{16})$ \label{tb1}} \\
No. & $3a_1$ & $\GNA$  \\ \hline \endfirsthead
\caption{$3V=(0^{16})$ (cont.)} \\
No. & $3a_1$ & $\GNA$ \\ \hline \endhead
1.1 & $(0^{16})$ & $SO(32)$  \\
1.2 & $(1^6,0^{10})$ & $SU(6) \ti SO(20)$ \\
1.3 & $(1^{12},0^{4})$ & $SU(12) \ti SO(8)$  \\
1.4 & $(-2,1^2,0^{13})$ & $SU(3) \ti SO(26)$  \\
1.5 & $(-2,1^8,0^{7})$ & $SU(9) \ti SO(14)$ \\
1.6 & $(-2,1^{14},0)$ & $SU(15)$ \\
\hline
\end{longtable}

\begin{longtable}{ccc}
\caption{$3V=(1^6;0^{10})$ \label{tb2}} \\
No. & $3a_1$ & $\GNA$  \\ \hline \endfirsthead
\caption{$3V=(1^6;0^{10})$ (cont.)} \\
No. & $3a_1$ & $\GNA$ \\ \hline \endhead
2.1 & $(0^{6};0^{10})$ & $SO(20)\ti SU(6)$ \\
2.2 & $(0^6;1^6,0^4)$ & $SO(8) \ti SU(6)^2$ \\
2.3 & $(0^6;-2,1^2,0^7)$ & $SO(14) \ti SU(6) \ti SU(3)$ \\
2.4 & $(0^6;-2,1^8,0)$ & $SU(9) \ti SU(6)$  \\
2.5 & $(1,-1,0^4;1^4,0^6)$ & $SO(12) \ti SU(4)^2$ \\
2.6, 2.7 & $(1,-1,0^4;1^9,\pm 1)$ & $SU(10) \ti SU(4)$  \\
2.8 & $(1^2,(-1)^2,0^2;1^2,0^8)$ & $SO(16) \ti SU(2)^4$  \\
2.9 & $(1^2,(-1)^2,0^2;1^8,0^2)$ & $SU(8) \ti SU(2)^5$  \\
2.10 & $(1^3,(-1)^3;0^{10})$ & $SO(20) \ti SU(3)^2$  \\
2.11 & $(1^3,(-1)^3;1^6,0^4)$ & $SU(6) \ti SO(8) \ti SU(3)^2$  \\
2.12 & $(1^3,0^3;1^3,0^7)$ & $SO(14) \ti SU(3)^3$ \\
2.13 & $(1^3,0^3;1^9,0)$ & $SU(9) \ti SU(3)^2$  \\
2.14 & $(1^4,-1,0;1,0^9)$ & $SO(18) \ti SU(4)$  \\
2.15 & $(1^4,-1,0;1^7,0^3)$ & $SU(7) \ti SU(4)^2$  \\
2.16 & $(1^6;0^{10})$ & $SO(20) \ti SU(6)$  \\
2.17 & $(1^6;1^6,0^4)$ & $SU(6)^2 \ti SO(8)$  \\
2.18 & $(-2,-1,0^4;1,0^9)$ & $SO(18) \ti SU(4)$  \\
2.19 & $(-2,-1,0^4;1^7,0^3)$ & $SU(7) \ti SU(4)^2$  \\
2.20 & $(-2,1,\mo^2,0^2;1^5,0^5)$ & $SO(10) \ti SU(5) \ti SU(2)^3$ \\
2.21 & $(-2,1^2,\mo^3;1^3,0^7)$ & $SO(14) \ti SU(3)^3$  \\
2.22 & $(-2,1^2,\mo^3;1^9,0)$ & $SU(9) \ti SU(3)^2$  \\
2.23 & $(-2,1^2,0^3;0^{10})$ & $SO(20) \ti SU(3)^2$  \\
2.24 & $(-2,1^2,0^3;1^6,0^4)$ & $SU(6) \ti SO(8) \ti SU(3)^2$  \\
2.25 & $(-2,1^3,-1,0;1^4,0^6)$ & $SO(12) \ti SU(4)^2$  \\
2.26, 2.27 & $(-2,1^3,-1,0;1^9,\pm 1)$ & $SU(10) \ti SU(4)$  \\
2.28 & $(-2,1^5;1^3,0^7)$ & $SO(14) \ti SU(6) \ti SU(3)$  \\
2.29 & $(-2,1^5;1^9,0)$ & $SU(9) \ti SU(6)$ \\
\hline
\end{longtable}

\begin{longtable}{ccc}
\caption{$3V=(1^{12};0^{4})$ \label{tb3}} \\
No. & $3a_1$ & $\GNA$  \\ \hline \endfirsthead
\caption{$3V=(1^{12};0^{4})$ (cont.)} \\
No. & $3a_1$ & $\GNA$ \\ \hline \endhead
3.1 & $(0^{12};0^{4})$          & $SU(12) \ti SO(8)$ \\
3.2 & $(0^{12};-2,1^2,0)$       & $SU(12) \ti SU(3)$ \\
3.3, 3.4 & $(1,-1,0^{10};1^3,\pm 1)$ & $SU(10) \ti SU(4)$ \\
3.5 & $(1^2,\mo^2,0^8;1^2,0^2)$ & $SU(8) \ti SU(2)^5$ \\
3.6 & $(1^3,\mo^3,0^6;0^4)$     & $SU(6) \ti SO(8) \ti SU(3)^2$ \\
3.7, 3.8 & $(1^4,\mo^4,0^4;1^3,\pm 1)$ & $SU(4)^4$ \\
3.9 & $(1^5,\mo^5,0^2;1^2,0^2)$ & $SU(5)^2 \ti SU(2)^4$ \\
3.10 & $(1^6,\mo^6;0^4)$         & $SU(6)^2 \ti SO(8)$ \\
3.11 & $(1^3,0^9;1^3,0)$        & $SU(9) \ti SU(3)^2$ \\
3.12 & $(1^4,-1,0^7;1,0^3)$     & $SU(7) \ti SU(4)^2$ \\
3.13 & $(1^6,\mo^3,0^3;1^3,0)$  & $SU(6) \ti SU(3)^3$ \\
3.14 & $(1^7,\mo^4,0;1,0^3)$    & $SU(7) \ti SU(4)^2$ \\
3.15 & $(1^6,0^6;0^4)$          & $SU(6)^2 \ti SO(8)$\\
3.16, 3.17 & $(1^7,-1,0^4;1^3,\pm 1)$ & $SU(7) \ti SU(4)^2$ \\
3.18 & $(1^8,\mo^2,0^2;1^2,0^2)$ & $SU(8) \ti SU(2)^5$ \\
3.19 & $(1^9,\mo^3;0^4)$        & $SU(9) \ti SO(8) \ti SU(3)$ \\
3.20 & $(1^9,0^3;1^3,0)$        & $SU(9) \ti SU(3)^2$ \\
3.21 & $(1^{10},-1,0;1,0^3)$    & $SU(10) \ti SU(4)$ \\
3.22 & $(1^{12};0^4)$           & $SU(12) \ti SO(8)$ \\
3.23 & $(-2,-1,0^{10};1,0^3)$   & $SU(10) \ti SU(4)$ \\
3.24 & $(-2,1^2,\mo^3,0^6;1^3,0)$ & $SU(6) \ti SU(3)^3$ \\
3.25 & $(-2,1^3,\mo^4,0^4;1,0^3)$ & $SU(4)^4$ \\
3.26 & $(-2,1^5,\mo^6;1^3,0)$   & $SU(6)^2 \ti SU(3)$ \\
3.27 & $(-2,1^2,0^9;0^4)$       & $SU(9) \ti SO(8) \ti SU(3)$ \\
3.28, 3.29 & $(-2,1^3,-1,0^7;1^3,\pm 1)$ & $SU(7) \ti SU(4)^2$ \\
3.30 & $(-2,1^4,\mo^2,0^5;1^2,0^2)$ & $SU(5)^2 \ti SU(2)^4$ \\
3.31 & $(-2,1^5,\mo^3,0^3;0^4)$ & $SU(6) \ti SO(8) \ti SU(3)^2$ \\
3.32, 3.33 & $(-2,1^6,\mo^4,0;1^3,\pm 1)$ & $SU(7) \ti SU(4)^2$ \\
3.34 & $(-2,1^5,0^6;1^3,0)$     & $SU(6)^2 \ti SU(3)$ \\
3.35 & $(-2,1^6,-1,0^4;1,0^3)$  & $SU(7) \ti SU(4)^2$ \\
3.36 & $(-2,1^8,\mo^3;1^3,0)$   & $SU(9) \ti SU(3)^2$ \\
3.37 & $(-2,1^8,0^3;0^4)$       & $SU(9) \ti SO(8) \ti SU(3)$ \\
3.38, 3.39 & $(-2,1^9,-1,0;1^3,\pm 1)$ & $SU(10) \ti SU(4)$ \\
3.40 & $(-2,1^{11};1^3,0)$      & $SU(12) \ti SU(3)$ \\
\hline
\end{longtable}

\begin{longtable}{ccc}
\caption{$3V=(-2,1^2;0^{13})$ \label{tb4}} \\
No. & $3a_1$ & $\GNA$  \\ \hline \endfirsthead
\caption{$3V=(-2,1^2;0^{13})$ (cont.)} \\
No. & $3a_1$ & $\GNA$ \\ \hline \endhead
4.1 & $(0^3;0^{13})$ & $SO(26) \ti SU(3)$ \\
4.2 & $(0^3;1^6,0^7)$ & $SO(14) \ti SU(6) \ti SU(3)$ \\
4.3 & $(0^3;1^{12},0)$ & $SU(12) \ti SU(3)$ \\
4.4 & $(0^3;-2,1^2,0^{10})$ & $SO(20) \ti SU(3)^2$ \\
4.5 & $(0^3;-2,1^8,0^4)$ & $SU(9) \ti SO(8) \ti SU(3)$ \\
4.6 & $(1,-1,0;1^4,0^9)$ & $SO(18) \ti SU(4)$ \\
4.7 & $(1,-1,0;1^{10},0^3)$ & $SU(10) \ti SU(4)$ \\
4.8 & $(1^3;1^3,0^{10})$ & $SO(20) \ti SU(3)^2$ \\
4.9 & $(1^3;1^9,0^4)$ & $SU(9) \ti SO(8) \ti SU(3)$ \\
4.10 & $(-2,-1,0;1,0^{12})$ & $SO(24)$ \\
4.11 & $(-2,-1,0;1^7,0^6)$ & $SO(12) \ti SU(7)$ \\
4.12, 4.13 & $(-2,-1,0;1^{12},\pm 1)$ & $SU(13)$ \\
4.14 & $(-2,1^2;0^{13})$ & $SO(26) \ti SU(3)$ \\
4.15 & $(-2,1^2;1^6,0^7)$ & $SO(14) \ti SU(6) \ti SU(3)$ \\
4.16 & $(-2,1^2;1^{12},0)$ & $SU(12) \ti SU(3)$ \\
\hline
\end{longtable}

\begin{longtable}{ccc}
\caption{$3V=(-2,1^8;0^7)$ \label{tb5}} \\
No. & $3a_1$ & $\GNA$  \\ \hline \endfirsthead
\caption{$3V=(-2,1^8;0^7)$ (cont.)} \\
No. & $3a_1$ & $\GNA$ \\ \hline \endhead
5.1 & $(0^9;0^7)$ & $SU(9) \ti SO(14)$ \\
5.2 & $(0^9;1^6,0)$ & $SU(9) \ti SU(6)$ \\
5.3 & $(0^9;-2,1^2,0^4)$ & $SU(9) \ti SO(8) \ti SU(3)$  \\
5.4 & $(1,-1,0^7;1^4,0^3)$ & $SU(7) \ti SU(4)^2$ \\
5.5 & $(1^2,\mo^2,0^5;1^2,0^5)$ & $SO(10) \ti SU(5) \ti SU(2)^3$ \\
5.6 & $(1^3,\mo^3,0^3;0^7)$ & $SO(14) \ti SU(3)^3$ \\
5.7 & $(1^3,\mo^3,0^3;1^6,0)$ & $SU(6) \ti SU(3)^3$ \\
5.8 & $(1^4,\mo^4,0;1^4,0^3)$ & $SU(4)^4$ \\
5.9 & $(1^3,0^6;1^3,0^4)$ & $SU(6) \ti SO(8) \ti SU(3)^2$ \\
5.10 & $(1^4,-1,0^4;1,0^6)$ & $SO(12) \ti SU(4)^2$ \\
5.11, 5.12 & $(1^4,-1,0^4;1^6,\pm 1)$ & $SU(7) \ti SU(4)^2$ \\
5.13 & $(1^5,\mo^2,0^2;1^5,0^2)$ & $SU(5)^2 \ti SU(2)^4$ \\
5.14 & $(1^6,\mo^3;1^3,0^4)$ & $SU(6) \ti SO(8) \ti SU(3)^2$ \\
5.15 & $(1^6,0^3;0^7)$ & $SO(14) \ti SU(6) \ti SU(3)$ \\
5.16 & $(1^6,0^3;1^6,0)$ & $SU(6)^2 \ti SU(3)$ \\
5.17 & $(1^7,-1,0;1^4,0^3)$ & $SU(7) \ti SU(4)^2$ \\
5.18 & $(1^9;1^3,0^4)$ & $SU(9) \ti SO(8) \ti SU(3)$ \\
5.19 & $(-2,-1,0^7;1,0^6)$ & $SO(12) \ti SU(7)$ \\
5.20, 5.21 & $(-2,-1,0^7;1^6,\pm 1)$ & $SU(7)^2$ \\
5.22 & $(-2,1,\mo^2,0^5;1^5,0^2)$ & $SU(5)^2 \ti SU(2)^4$ \\
5.23 & $(-2,1^2,\mo^3,0^3;1^3,0^4)$ & $SO(8) \ti SU(3)^4$ \\
5.24 & $(-2,1^3,\mo^4,0;1,0^6)$ & $SO(12) \ti SU(4)^2$ \\
5.25, 5.26 & $(-2,1^3,\mo^4,0;1^6,\pm 1)$ & $SU(7) \ti SU(4)^2$ \\
5.27 & $(-2,1^2,0^6;0^7)$ & $SO(14) \ti SU(6) \ti SU(3)$ \\
5.28 & $(-2,1^2,0^6;1^6,0)$ & $SU(6)^2 \ti SU(3)$ \\
5.29 & $(-2,1^3,-1,0^4;1^4,0^3)$ & $SU(4)^4$ \\
5.30 & $(-2,1^4,\mo^2,0^2;1^2,0^5)$ & $SO(10) \ti SU(5) \ti SU(2)^3$ \\
5.31 & $(-2,1^5,\mo^3;0^7)$ & $SO(14) \ti SU(6) \ti SU(3)$ \\
5.32 & $(-2,1^5,\mo^3;1^6,0)$ & $SU(6)^2 \ti SU(3)$ \\
5.33 & $(-2,1^5,0^3;1^3,0^4)$ & $SU(6) \ti SO(8) \ti SU(3)^2$ \\
5.34 & $(-2,1^6,-1,0;1,0^6)$ & $SO(12) \ti SU(7)$ \\
5.35, 5.36 & $(-2,1^6,-1,0;1^6,\pm 1)$ & $SU(7)^2$ \\
5.37 & $(-2,1^8;0^7)$ & $SU(9) \ti SO(14)$ \\
5.38 & $(-2,1^8;1^6,0)$ & $SU(9) \ti SU(6)$ \\
\hline
\end{longtable}

\begin{longtable}{ccc}
\caption{$3V=(-2,1^{14},0)$ \label{tb6}} \\
No. & $3a_1$ & $\GNA$  \\ \hline \endfirsthead
\caption{$3V=(-2,1^{14};0)$ (cont.)} \\
No. & $3a_1$ & $\GNA$ \\ \hline \endhead
6.1 & $(0^{15};0)$ & $SU(15)$ \\
6.2 & $(1^3,\mo^3,0^9;0)$ & $SU(9) \ti SU(3)^2$ \\
6.3 & $(1^6,\mo^6,0^3;0)$ & $SU(6)^2 \ti SU(3)$ \\
6.4, 6.5 & $(1^4,-1,0^{10};\pm 1)$ & $SU(10) \ti SU(4)$ \\
6.6, 6.7 & $(1^7,\mo^4,0^4;\pm 1)$ & $SU(7) \ti SU(4)^2$ \\
6.8 & $(1^6,0^9;0)$ & $SU(9) \ti SU(6)$ \\
6.9 & $(1^9,\mo^3,0^3;0)$ & $SU(9) \ti SU(3)^2$  \\
6.10, 6.11 & $(1^{10},-1,0^4;\pm 1)$ & $SU(10) \ti SU(4)$  \\
6.12 & $(1^{12},0^3;0)$ & $SU(12) \ti SU(3)$ \\
6.13, 6.14 & $(-2,-1,0^{13};\pm 1)$ & $SU(13)$ \\
6.15, 6.16 & $(-2,1^3,\mo^4,0^7;\pm 1)$ & $SU(7) \ti SU(4)^2$  \\
6.17, 6.18 & $(-2,1^6,\mo^7,0;\pm 1)$ & $SU(7)^2$ \\
6.19 & $(-2,1^2,0^{12};0)$ & $SU(12) \ti SU(3)$  \\
6.20 & $(-2,1^5,\mo^3,0^6;0)$ & $SU(6)^2 \ti SU(3)$  \\
6.21 & $(-2,1^8,\mo^6;0)$ & $SU(9) \ti SU(6)$  \\
6.22, 6.23 & $(-2,1^6,-1,0^7;\pm 1)$ & $SU(7)^2$  \\
6.24, 6.25 & $(-2,1^9,\mo^4,0;\pm 1)$ & $SU(10) \ti SU(4)$  \\
6.26 & $(-2,1^8,0^6;0)$ & $SU(9) \ti SU(6)$  \\
6.27 & $(-2,1^{11},\mo^3;0)$ & $SU(12) \ti SU(3)$  \\
6.28, 6.29 & $(-2,1^{12},-1,0;\pm 1)$ & $SU(13)$  \\
6.30 & $(-2,1^{14};0)$ & $SU(15)$ \\
\hline
\end{longtable}


\newpage

\begin{thebibliography}{99}

\bibitem{GHMR85}
D. J. Gross, J. A. Harvey, E. Martinec and R. Rohm,
Phys. Rev. Lett. 54 (1985) 502.

\bibitem{gcd}
H.-P. Nilles,
Phys. Lett. B 115 (1982) 193;
S. Ferrara, L. Girardello, H.-P. Nilles,
Phys. Lett. B 125 (1983) 457.

\bibitem{DHVW85}
L. Dixon, J. Harvey, C. Vafa and E. Witten,
Nucl. Phys. B 261 (1985) 678.

\bibitem{DHVW86}
L. Dixon, J. Harvey, C. Vafa and E. Witten,
Nucl. Phys. B 274 (1986) 285.

\bibitem{Gie02a}
J. Giedt,
Ann. of Phys. (N.Y.) 297 (2002) 67\xxx{hep-th/0108244}.

\bibitem{flc}
A. Font, L. E. Ib\'a\~nez, F. Quevedo and A. Sierra,
Nucl. Phys. B 331 (1990) 421;
J. A. Casas and C. Mu\~noz,
Phys. Lett. B 209 (1988) 214;
B 214 (1988) 63;
G. B. Cleaver, A. E. Faraggi and D. V. Nanopoulos,
Phys. Lett. B 455 (1999) 135;
Int. J. Mod. Phys. A 16 (2001) 425;
G. B. Cleaver, A. E. Faraggi, D. V. Nanopoulos, and
J. W. Walker, Mod. Phys. Lett. A 15 (2000) 1191;
Nucl. Phys. B 593 (2001) 471; hep-ph/0104091.

\bibitem{FIU}
M. B. Green and J. H. Schwarz,
Phys. Lett. B 149 (1984) 117;
M. Dine, N. Seiberg and E. Witten,
Nucl. Phys. B 289 (1987) 585;
J. J. Atick, L. Dixon and A. Sen,
Nucl. Phys. B 292 (1987) 109;
M. Dine, I. Ichinose and N. Seiberg,
Nucl. Phys. B 293 (1987) 253.

\bibitem{dmstab}
M. K. Gaillard, J. Giedt,
Phys. Lett. B 479 (2000) 308\xxx{hep-ph/0001219};
Nucl. Phys. B 636 (2002) 365\xxx{hep-th/0204100};
Nucl. Phys. B 643 (2002) 201\xxx{hep-th/0206249};
J. Giedt, Mod. Phys. Lett. A 17 (2002) 1465\xxx{hep-ph/0204017};
hep-ph/0208004.

\bibitem{GM}
G.F. Giudice, R. Rattazzi,
Phys. Rept. 322 (1999) 419\xxx{hep-ph/9801271}.

\bibitem{IMNQ}
L. E. Ib\'a\~nez, H.-P. Nilles and F. Quevedo,
Phys. Lett. B 187 (1987) 25;
L. E. Ib\'a\~nez, J. Mas, H.-P. Nilles and F. Quevedo,
Nucl. Phys. B 301 (1988) 157.

\bibitem{CMM89}
J. A. Casas, M. Mondragon and C. Mu\~noz,
Phys. Lett. B 230 (1989) 63.

\bibitem{Gie01b}
J. Giedt,
Ann. of Phys. (N.Y.) 289 (2001) 251\xxx{hep-th/0009104}.

\bibitem{AFIU95}
G. Aldazabal, A. Font, L. E. Ib\'a\~nez, A. M. Uranga,
Nucl. Phys. B 452 (1995) 3\xxx{hep-th/9410206};
Nucl. Phys. B 465 (1996) 34 (1995) \xxx{hep-th/9508033}.

\bibitem{sgut:o}
D. C. Lewellen, Nucl. Phys. B 337 (1990) 61;
A. Font, L. Ib\'a\~nez, F. Quevedo,
Nucl. Phys. B 345 (1990) 389;
G. B. Cleaver, hep-th/9409096; hep-th/9506006; hep-th/9604183;
Chaudhuri, S. W. Chung and J. D. Lykken,
hep-ph/9405374;
S. Chaudhuri, S. W. Chung, G. Hockney, J. Lykken,
hep-th/9409151;
Nucl. Phys. B 456 (1995) 89\xxx{hep-ph/9501361}.

\bibitem{sgut:d}
K. R. Dienes, J. March-Russell,
Nucl. Phys. B 479 (1996) 113\xxx{hep-th/9604112};
K. R. Dienes,
Nucl.Phys. B 488 (1997) 141\xxx{hep-ph/9606467}.

\bibitem{sgut:kt}
Z. Kakushadze and S. H. Tye,
Phys. Rev. Lett. 77 (1996) 2612\xxx{hep-th/9605221};
Phys. Rev. D 54 (1996) 7520\xxx{hep-th/9607138};
Phys. Lett. B 392 (1997) 335\xxx{hep-th/9609027};
Phys. Rev. D 55 (1997) 7878\xxx{hep-th/9610106};
Phys. Rev. D 55 (1997) 7896\xxx{hep-th/9701057};
Z. Kakushadze, G. Shiu, S. H. Tye,
Phys. Rev. D 54 (1996) 7545\xxx{hep-th/9607137};
Nucl. Phys. B 501 (1997) 547\xxx{hep-th/9704113};
Z. Kakushadze, G. Shiu, S. H. Tye, Y. Vtorov-Karevsky,
Int.  J.  Mod.  Phys.  A 13 (1998) 2551\xxx{hep-th/9710149};
Phys. Lett. B 408 (1997) 173\xxx{hep-ph/9705202}.

\bibitem{Kim03}
J. E. Kim, hep-th/0301177.

\end{thebibliography}
\end{document}